\documentclass[journal=ancac3,manuscript=article]{achemso}
\setkeys{acs}{articletitle=true,etalmode=truncate}

\usepackage[version=3]{mhchem} % Formula subscripts using \ce{}
\usepackage[T1]{fontenc}       % Use modern font encodings
\usepackage{graphicx}
\usepackage{color}
\usepackage{float}
\usepackage[normalem]{ulem}

\author{Anike Purbawati}
\affiliation{Univ. Grenoble Alpes, CNRS, Grenoble INP, Institut NEEL, 38000 Grenoble, France}

\author{Johann Coraux}
\affiliation{Univ. Grenoble Alpes, CNRS, Grenoble INP, Institut NEEL, 38000 Grenoble, France}
\email{johann.coraux@neel.cnrs.fr}

\author{Jan Vogel}
\affiliation{Univ. Grenoble Alpes, CNRS, Grenoble INP, Institut NEEL, 38000 Grenoble, France}

\author{Abdellali Hadj-Azzem}
\affiliation{Univ. Grenoble Alpes, CNRS, Grenoble INP, Institut NEEL, 38000 Grenoble, France}

\author{NianJheng Wu}
\affiliation{Univ. Grenoble Alpes, CNRS, Grenoble INP, Institut NEEL, 38000 Grenoble, France}

\author{Nedjma Bendiab}
\affiliation{Univ. Grenoble Alpes, CNRS, Grenoble INP, Institut NEEL, 38000 Grenoble, France}

\author{David Jegouso}
\affiliation{Univ. Grenoble Alpes, CNRS, Grenoble INP, Institut NEEL, 38000 Grenoble, France}

\author{Julien Renard}
\affiliation{Univ. Grenoble Alpes, CNRS, Grenoble INP, Institut NEEL, 38000 Grenoble, France}

\author{Laetitia Marty}
\affiliation{Univ. Grenoble Alpes, CNRS, Grenoble INP, Institut NEEL, 38000 Grenoble, France}

\author{Vincent Bouchiat}
\affiliation{Univ. Grenoble Alpes, CNRS, Grenoble INP, Institut NEEL, 38000 Grenoble, France}

\author{Andr\'e Sulpice}
\affiliation{Univ. Grenoble Alpes, CNRS, Grenoble INP, Institut NEEL, 38000 Grenoble, France}

\author{Lucia Aballe}
\affiliation{ALBA Synchrotron Light Facility, 08290 Cerdanyola del Valles, Spain}

\author{Michael Foerster}
\affiliation{ALBA Synchrotron Light Facility, 08290 Cerdanyola del Valles, Spain}

\author{Francesca Genuzio}
\affiliation{Elettra-Sincrotrone Trieste S.C.p.A., S:S. 14, km 163.5 in AREA Science Park, Basovizza, 34149 Trieste, Italy}

\author{Andrea Locatelli}
\affiliation{Elettra-Sincrotrone Trieste S.C.p.A., S:S. 14, km 163.5 in AREA Science Park, Basovizza, 34149 Trieste, Italy}

\author{Tevfik Onur Mente\c{s}}
\affiliation{Elettra-Sincrotrone Trieste S.C.p.A., S:S. 14, km 163.5 in AREA Science Park, Basovizza, 34149 Trieste, Italy}

\author{Zheng Vitto Han}
\affiliation{Shenyang National Laboratory for Materials Science, Institute of Metal Research, Chinese Academy of Sciences, Shenyang 110016, China}
\alsoaffiliation{School of Material Science and Engineering, University of Science and Technology of China, Anhui 230026, China}
\alsoaffiliation{Collaborative Innovation Center of Extreme Optics, Shanxi University, Taiyuan 030006, P.R.China}

\author{Xingdan Sun}
\affiliation{Shenyang National Laboratory for Materials Science, Institute of Metal Research, Chinese Academy of Sciences, Shenyang 110016, China}
\alsoaffiliation{School of Material Science and Engineering, University of Science and Technology of China, Anhui 230026, China}

\author{Manuel N\'u\~nez-Regueiro}
\affiliation{Univ. Grenoble Alpes, CNRS, Grenoble INP, Institut NEEL, 38000 Grenoble, France}

\author{Nicolas Rougemaille}
\affiliation{Univ. Grenoble Alpes, CNRS, Grenoble INP, Institut NEEL, 38000 Grenoble, France}

\title[Ferromagnetic CrTe$_2$ flakes]
  {In-plane magnetic domains and N\'eel-like domain walls in thin flakes of the room temperature CrTe$_2$ van der Waals ferromagnet}

\begin{document}


\bigskip
\bigskip

\begin{abstract}
The recent discovery of magnetic van der Waals materials has triggered a wealth of investigations in materials science, and now offers genuinely new prospects for both fundamental and applied research.
Although the catalogue of van der Waals ferromagnets is rapidly expanding, most of them have a Curie temperature below 300~K, a notable disadvantage for potential applications.
Combining element-selective x-ray magnetic imaging and magnetic force microscopy, we resolve at room temperature the magnetic domains and domains walls in micron-sized flakes of the CrTe$_2$ van der Waals ferromagnet.
Flux-closure magnetic patterns suggesting in-plane six-fold symmetry are observed.
Upon annealing the material above its Curie point (315~K), the magnetic domains disappear. 
By cooling back down the sample, a different magnetic domain distribution is obtained, indicating material stability and lack of magnetic memory upon thermal cycling.
The domain walls presumably have N\'eel texture, are preferentially oriented along directions separated by 120 degrees, and have a width of several tens of nanometers.
Besides microscopic mapping of magnetic domains and domain walls, the coercivity of the material is found to be of a few mT only, showing that the CrTe$_2$ compound is magnetically soft. The coercivity is found to increase as the volume of the material decreases.
\end{abstract}


%A few useful instructions in ACS style:
%Supporting Information
%Materials and Methods
%Figure
%\citenum{ref}

\newpage

\section*{Introduction}

Magnetic van der Waals (vdW) materials prove to be an excellent playground in which to explore magnetic phenomena in low dimensions, and are candidate building blocks for spintronic devices.
On the one hand, the origin of the magnetic order in vdW ferromagnets is often scrutinized under the prism of canonical spin lattice models, seeking for signatures of two-dimensional exotic magnetic behaviors.\cite{Gong2017,Huang2017,Deng2018,Fei2018,Burch2018,Wong2019,Gibertini2019}
On the other, applied research is devising new strategies to tune their magnetic properties through electrostatic control, opening new avenues for applications in spintronics\cite{Xing2017,Deng2018,Jiang2018,Jiang2018b,Wang2018a}. 
A few prototypical devices have been already demonstrated, including magnetic tunnel junctions with large magneto-resistance\cite{Kim2018,Klein2018,Song2018,Wang2018b,Kim2019,Albarakati2019} . 
Although the catalogue of vdW ferromagnets is rich and extends well beyond the most studied CrI$_3$ and Fe$_3$GeTe$_2$ compounds, the palette of available magnetic properties remains partial.
In that context, computational efforts may guide the search for materials with specific properties,\cite{Jiang2018c,Torelli2019} especially those with a high Curie temperature (T$_\text{C}$). 

Measurements of high T$_\text{C}$ were reported for three few- and even single-layer dichalcogenides with 1T polytype\cite{note1}, VSe$_2$, VTe$_2$, and MnTe$_2$\cite{Bonilla2018,Li2018,Yu2019,Ohara2018}, although contradicting results are reported in some cases.\cite{Feng2018,Wong2019} 
A near room temperature T$_\mathrm{C}$ was also reported in Fe$_5$GeTe$_2$,\cite{May2019} and in Fe$_3$GeTe$_2$ ultra-thin films,\cite{Deng2018} the latter under the influence of ionic gating (the chemical composition of Fe$_3$GeTe$_2$ being altered by the gating process itself\cite{Weber2019}).
Further, the Curie temperature can be substantially increased (above 300~K) in 250~nm-thick Fe$_3$GeTe$_2$ flakes etched in the form of microstructures.\cite{Li2018b}
Interestingly, this microstructuration also induces an in-plane magnetization,\cite{Li2018b} a rare situation in ultra-thin vdW ferromagnets. 
Beyond this particular case, the CrCl$_3$ compound has an intrinsic in-plane magnetization, down to few atomic layers, but with a Curie temperature well below 300~K.\cite{Cai2019} 
To date, there is no known vdW ferromagnet with thickness below 100~nm, and with an in-plane magnetization and a T$_\text{C}$ above room temperature.
Ferromagnetism above room temperature would offer new opportunities to study spintronic devices.
In-plane magnetization for an ultra-thin layer would provide further possibilities, for instance to test remarkable phase transitions in two-dimensional XY spin lattice models,\cite{Kosterlitz1973} to stabilise quantum phases via proximity effects,\cite{Liu2013} and to build synthetic antiferromagnets with tunable magnetic properties.\cite{Duine2018} 

Here, we demonstrate both ferromagnetism above room temperature and in-plane magnetic anisotropy in exfoliated flakes of the 1T-CrTe$_2$ vdW transition metal dichalcogenide, an established ferromagnet in its bulk form.\cite{Freitas2015} 
Information on the lateral distribution of magnetic domains is often missing in considering the properties of thin vdW ferromagnets.\cite{Tan2018,Zhang2019}
Therefore, we use two magnetic imaging techniques, magnetic force microscopy (MFM) and x-ray magnetic circular dichroism photoemission electron microscopy (XMCD-PEEM) to unveil the structure of magnetic domains and their domain walls. 
We unveil flux-closure in-plane magnetic domains, uniform across the thickness of the flakes, and separated by N\'{e}el domain walls at the surface of the flake, typically extending across 50 nm. 
Complementing our microscopy measurements by spatially-resolved Kerr magnetometry, we find a close-to-full remanence that cannot be observed in bulk crystals for magnetostatic reasons. Besides, we observe that the coercivity is small and depends on the volume of the flake. 
Finally, based on Raman spectroscopy, x-ray absorption spectroscopy (XAS) and x-ray photoelectron spectroscopy (XPS), we address the stability of the exfoliated CrTe$_2$ compound in ambient conditions, and find that an oxidation process takes place on a time scale of about a day.

\section*{Results and Discussion}

\begin{figure}[hbt]
 \begin{center}
 \includegraphics[width=8cm]{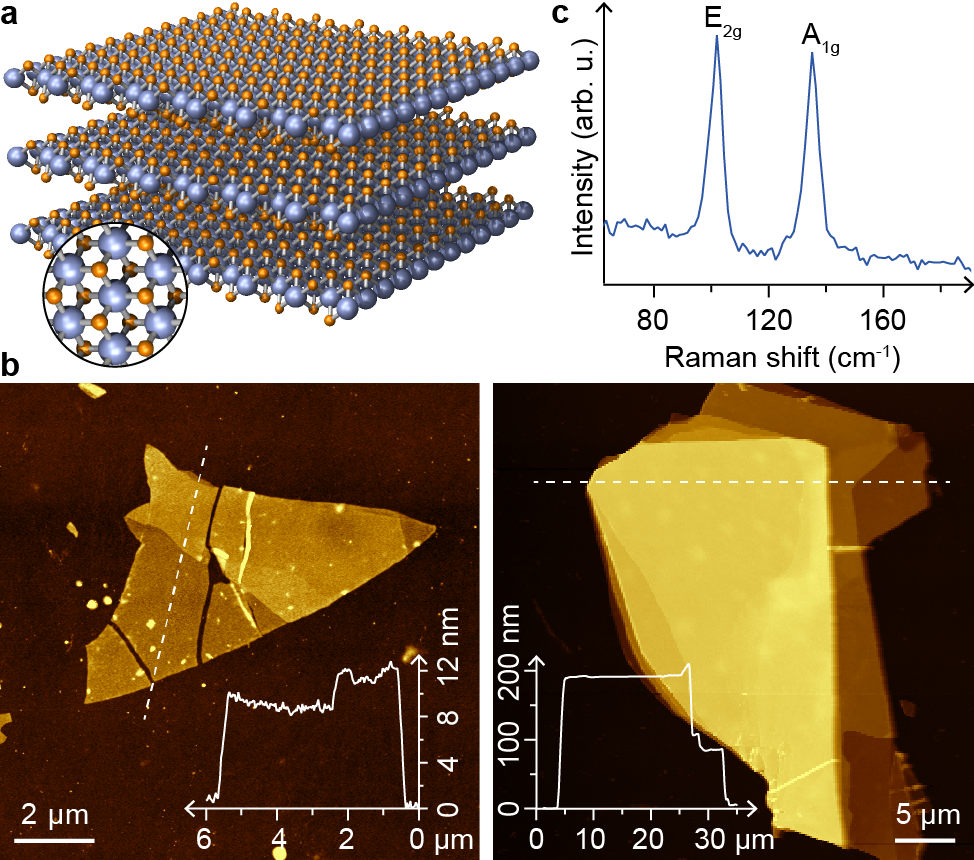}
 \caption{\label{fig1} \textbf{Structure, morphology and Raman reference spectra of exfoliated CrTe$_2$.} (a) Crystal structure of bulk CrTe$_2$. (b) AFM topographs of exfoliated CrTe$_2$ flakes with a thickness of the order of 10~nm (left) and 200~nm (right). The superimposed curves show a height profile of the two flakes. (c) Raman spectrum of an exfoliated, 13-nm-thick flake prepared under inert conditions revealing two main peaks (E$_{2\text{g}}$, A$_{1\text{g}}$) at 102 and 134 cm$^{-1}$.}
 \end{center}
\end{figure}

\subsection*{Exfoliation of thin CrTe$_2$ flakes} Bulk CrTe$_2$ was synthesized in the 1T structure (Figure~\ref{fig1}a), starting from the removal of potassium in a bulk KCrTe$_2$ crystal (see \textit{Materials and Methods}), following a procedure described elsewhere.\cite{Freitas2015} To prepare thin flakes of CrTe$_2$, we tested several mechanical exfoliation processes, using a standard scotch tape, a viscoelastic stamp of polydimethylsiloxane, and a thermal release tape. Whatever the process, the yield of thin (\textit{i.e.} thinner than 20~nm) flakes proved to be much lower than in the case of other, better-known vdW materials such as graphene or MoS$_2$. The lateral extension of the flakes was found to decrease with the flake thickness, being of the order of 10~$\mu$m for $\sim$10~nm-thick flakes, as assessed with atomic force microscopy (AFM, Figure~\ref{fig1}b).
The flakes were transferred onto SiO$_2$(285 nm)/Si and Pt(1 nm)/Ta(10 nm)/Si host substrates. In AFM images, we often observe nanoclusters covering a small fraction of the surface (see Figure~\ref{fig1}b). These clusters exhibit a slight specific phase contrast in AFM (few degrees difference) compared to the clean flake surface, indicative that they slightly modify mechanical dissipation processes. In our local XAS and XPS experiments, they exhibit no particular contrast. We tentatively ascribe them as dust particles, or scotch tape residues generated by the exfoliation process, that are weakly bonded to the surface.

The Raman spectra of the freshly exfoliated flakes on SiO$_2$/Si feature two characteristic bands (Figure~\ref{fig1}c), the E$_{2\text{g}}$ and A$_{1\text{g}}$ modes expected for 1T-CrTe$_2$ (Figure~S1).
In the thickness range we probe in this work ($\sim$10~nm to $\sim$200 nm), the wavenumbers (102 and 134~cm$^{-1}$) of these two bands are constant, unlike their intensity, which is weaker for thicker flakes  (see Supporting Information, Figure~S1). 
This thickness dependence of the intensity is attributed to the metallic character of 1T-CrTe$_2$:\cite{Freitas2015} while optical interferences within the gap defined by the Si surface and the bottom CrTe$_2$ surface are expected to enhance the Raman signal for a SiO$_2$ thickness of the order of the optical wavelength,\cite{Buscema2014,Zhang2015} the penetration depth of light in metallic CrTe$_2$ is presumably of the order of few/several 10~nm.
Under such conditions, interference effects have a vanishing effect as the flake becomes thicker than several tens of nm.

\begin{figure}[hbt]
 \begin{center}
 \includegraphics[width=11cm]{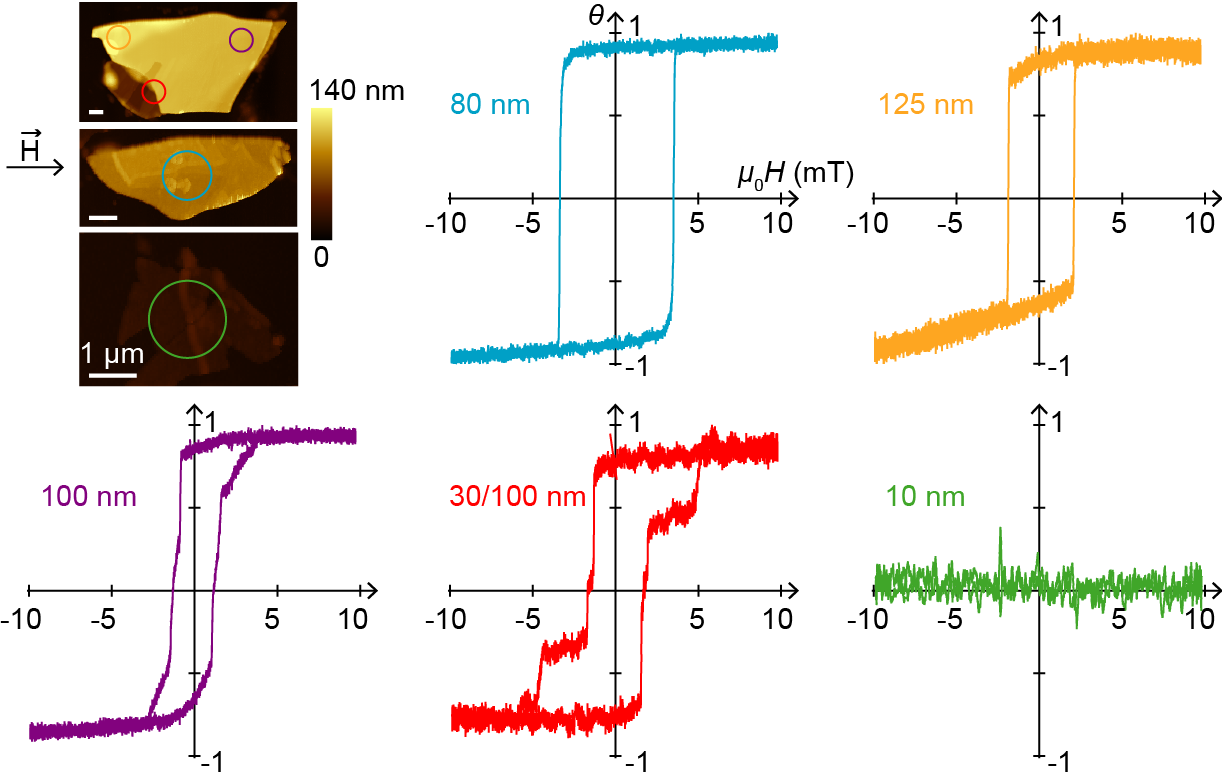}
 \caption{\label{fig2} \textbf{Spatially-resolved magnetometry.} Focused Kerr measurements performed in longitudinal geometry and at room temperature on three CrTe$_2$ flakes having different thicknesses, as indicated in the AFM image. On the thinnest flake (10~nm, green), no magnetic signal is found, whereas a Kerr signal is clearly seen for flakes with thickness of 30 (red), 80 (blue\cite{noteondust}), 100 (purple and red) and 125~nm (yellow). In all cases, the hysteresis loops show strong remanence, close to 100\%, and a coercivity of a few mT only.}
 \end{center}
\end{figure}

\subsection*{Spatially-resolved magnetometry} The magnetic properties of exfoliated CrTe$_2$ flakes were first characterized at room temperature, using focused magneto-optical Kerr magnetometry (see \textit{Materials and Methods}). 
The technique allows recording hysteresis loops under a focused laser spot of a size smaller than 1~$\mu$m, and an external magnetic field $\mu_0H$ applied in the plane of the film.
Figure~\ref{fig2} shows examples of hysteresis loops measured on different locations, indicated by a colored circle in the AFM images. 
Three CrTe$_2$ flakes with different thicknesses are probed. 
Except for the thinnest flake (10~nm), a magnetic hysteresis loop is systematically observed, providing a direct evidence of room temperature ferromagnetism. 
The fact that no Kerr signal is measured for flakes thinner than 20~nm could be attributed to a reduced Curie temperature, to a limited sensitivity of our magneto-optical Kerr magnetometer, or to the degredation of the material surface under ambient conditions (see below).
The hysteresis loops are essentially square, showing a remanent magnetization close to 100\%. 
This result indicates that the CrTe$_2$ flakes are in-plane magnetized, and that magnetic domains should be larger than the laser spot ($\sim$1~$\mu$m). 
The coercive field $\mu_0 H_\mathrm{c}$ is of the order of a few mT, a relatively low value characteristic of a magnetically soft material.

The coercive field varies from flake to flake, and even within a single flake (Figure~\ref{fig2} and Figure~S2). To understand the origin of these variations, we first discuss the case of the orange and purple hysteresis loops, which were both acquired on the same flake. At this point it is important to remind what is occurring during the focussed Kerr measurements as function of the swept magnetic field. The Kerr rotation is measured within the narrow footprint of the focused laser beam, but this does not mean that the characterisation is local at this scale. Indeed, the swept external magnetic field (which is applied uniformly, at larger scale) induces magnetisation reversal starting with the nucleation, possibly out of the laser spot, of reversed domains surrounded by domain walls. Magnetisation reversal goes on \textit{via} the motion of these domain walls, and hence happens with a delay away from the reversed domain nucleation centers. During this delay the sweep of the magnetic field has continued. This means that locally, away from the nucleation centers, magnetisation reversal will effectively occur at high magnetic field, \textit{i.e.} the measured coercivity will be higher than at the nucleation centers. This is true if domain wall propagation can occur from the nucleation centers to the laser spot.

However, the two regions (orange and purple) are at least partly magnetically independent. To maintain the smoothness of the discussion, we refrain from disclosing the analysis of magnetisation patterns presented in the next section, but still need to mention that this independence is evident in the magnetic imaging data of the same flake (Figure~\ref{fig3}a, top pannel, and similar kind of data acquired with varying applied magnetic field): the top-left elongated triangle which corresponds to the thicker region in the top pannel of Figure~\ref{fig2} (where the orange hysteresis loop was acquired) and the remaining of the flake (where the purple hysteresis loop was acquired) exhibit seemingly unrelated magnetisation patterns.

Why these regions are independent is unclear based on our experimental data. We  suggest that the independence has its origin in the presence of a boundary between the two regions, within a more or less thick wall, where the structure or composition of the material is altered, hence ferromagnetism is suppressed (there, the material could for instance have a non magnetic 2H structure instead of the 1T one).

The region corresponding to the orange hysteresis loop is slightly thicker, but its area is much smaller than the area corresponding to the purple hysteresis loop. The former hence has smaller volume than the latter, and since coercivity is increased if the number of nucleation centres (local defects) for reversed magnetic domains is decreased,\cite{Livingston} we indeed expect fewer domain nucleation centres (if we assume a constant local defect concentration in CrTe$_2$) and higher coercivity for the orange hysteresis loop than for the purple hysteresis loop. We transposed this analysis to 10 other hysteresis loops. In some cases, the laser spot illuminates two regions that are magnetically independent, in which case the hysteresis loop exhibits a step-like reversal (\textit{e.g.} red hysteresis loop in Figure~\ref{fig2}). Overall, we found that the increase of coercivity with decreasing volume is a robust trend. It is interesting to note that lamellar materials such as CrTe$_2$ are naturally immune to an increase in their surface roughness with the flake thickness. This is in contrast to more traditional thin films grown from the bottom up, where roughness, increasing with the thickness, was argued to alter coercivity.\cite{Palasantzas}

We end up this section by noting the strong difference in the magnetisation measured with SQUID magnetometry \textit{versus} applied magnetic field, for thin flakes (Figure~\ref{fig2}) and for macroscopic flakes. No or very little remanence and loop opening is observed in the latter case, neither for in-plane nor for out-of-plane magnetic fields down to low temperatures (Figure~S3). The saturation magnetisation $M_\mathrm{S}$ is reached for lower in-plane than out-of-plane magnetic fields, which is indicative, here as well, of a preferential orientation of the magnetisation in the plane of the flakes. What is the origin of this preference?

From the SQUID data we deduce a relatively small $M_\mathrm{S}$ value of 250~kA/m (\textit{i.e.} slightly less than the third of the value for permalloy and a sixth of the value in cobalt). In the case of an ideal flat sheet, with a perpendicular demagnetisation factor of 1, $\mu_0 M_\mathrm{S}$ = 310~mT would be needed to saturate the magnetisation perpendicular to the surface. A more realistic $<$1 value of the demagnetisation factor is expected for the macroscopic CrTe$_2$ flakes, which have finite thickness and extension, hence a slightly lower field would be needed here. This field should however still be larger than the $\sim$50~mT value that can be inferred by determining the field at which a zero-field linear fit to the data reaches $M_\mathrm{S}$ (Figure~S3). This is an indication that shape anisotropy alone cannot account for the observed anisotropy of the SQUID data. Antagonist magnetocrystalline effects favouring out-of-plane magnetisation provide a reasonable interpretation.

\subsection*{In-plane, flux-closure magnetization patterns and magnetic domain walls} At first thought, the very weak or absence of remanence observed in the SQUID data corresponding to the macroscopic flakes may seem a contradiction with the 100\% remanence observed in the focussed Kerr data. However, we also expect globally, for each exfoliated thin flake, no loop opening, despite local loop openings at smaller scale (as observed in the focussed Kerr measurements). The reason is, as we will now see, that magnetic domains in the thin flakes form flux-closure patterns. To unveil these patterns, we used two complementary magnetic imaging techniques (see \textit{Materials and Methods}):
\begin{itemize}
\setlength\itemsep{0em}
\item Magnetic force microscopy (MFM), to probe the magnetic stray fields emitted by the magnetization texture, especially to detect magnetic domain walls.
\item Photoemission electron microscopy (PEEM) combined with x-ray magnetic circular dichroism (XMCD), to probe the local direction of magnetization.
\end{itemize}

\begin{figure}[!hbt]
 \begin{center}
 \includegraphics[width=10cm]{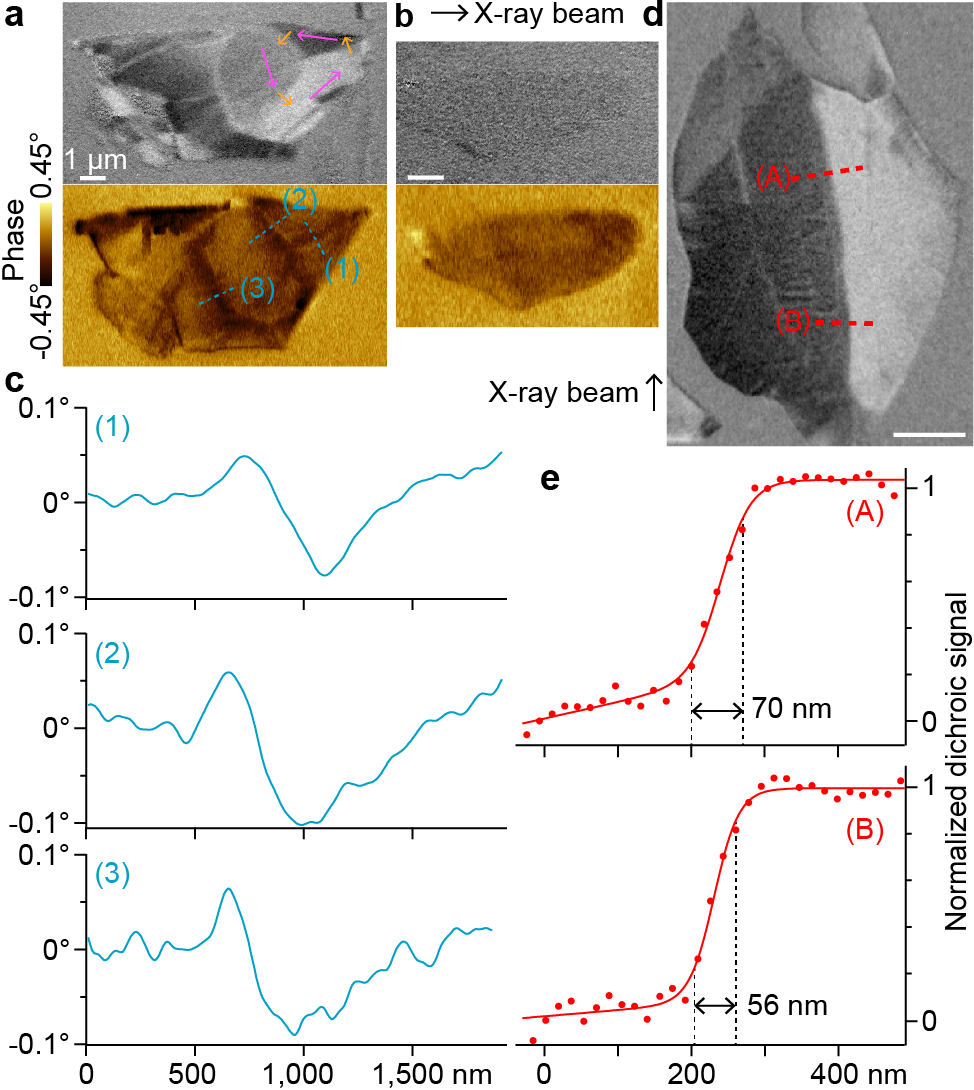}
 \caption{\label{fig3} \textbf{Magnetic texture of individual flakes - domains and domain walls.} (a,b) XMCD-PEEM (top panels) and MFM (bottom panels) images on the two flakes ((a) 30-100-125~nm-thick, (b) 80~nm-thick) reported in Figure~\ref{fig2} and exhibiting ferromagnetism at room temperature. The XMCD-PEEM images are obtained at the Cr L$_3$ absorption edge (575.8~eV). (c) Analysis of the magnetic domain walls in the MFM image shown in (a). The three intensity profiles are associated to the blue dashed lines un (a). (d) High-resolution XMCD-PEEM image of a 25 nm-thick flake. (e) Analysis of the magnetic domain wall width through the two red dashed lines shown in (d). Scale bar is 1 $\mu$m in all images. The direction of the x-ray beam is indicated with horizontal (a,b) and vertical (d) arrows.}
 \end{center}
\end{figure}

\noindent
The spatial resolution of the two techniques is similar, and of the order of a few tens of nanometers, typically. 
Contrary to MFM, XMCD-PEEM is surface sensitive, and only the top few nanometers below the material surface are imaged.
The two techniques were applied to a large number of CrTe$_2$ flakes, and in some cases to the very same flakes. 
Figures~\ref{fig3}a,b show the MFM and XMCD-PEEM magnetic contrasts obtained on the two magnetic flakes also studied with focused Kerr magnetometry and shown in Figure~\ref{fig2}.
The two images for the thickest flake (Figure~\ref{fig3}a) reveal a magnetic contrast at room temperature, and are obtained on an exfoliated sample kept in ambient conditions for several months. 
As we will see below, although the material's surface oxidizes with time, the magnetic texture of the flakes can be imaged reproducibly over an extended period of time. 
In addition, both techniques were used on the same magnetic configuration (no external magnetic field was applied between the MFM and XMCD-PEEM experiments), so that the magnetic images reported in Figures~\ref{fig3}a,b can be compared directly.

\begin{figure}[!hbt]
 \begin{center}
 \includegraphics[width=8cm]{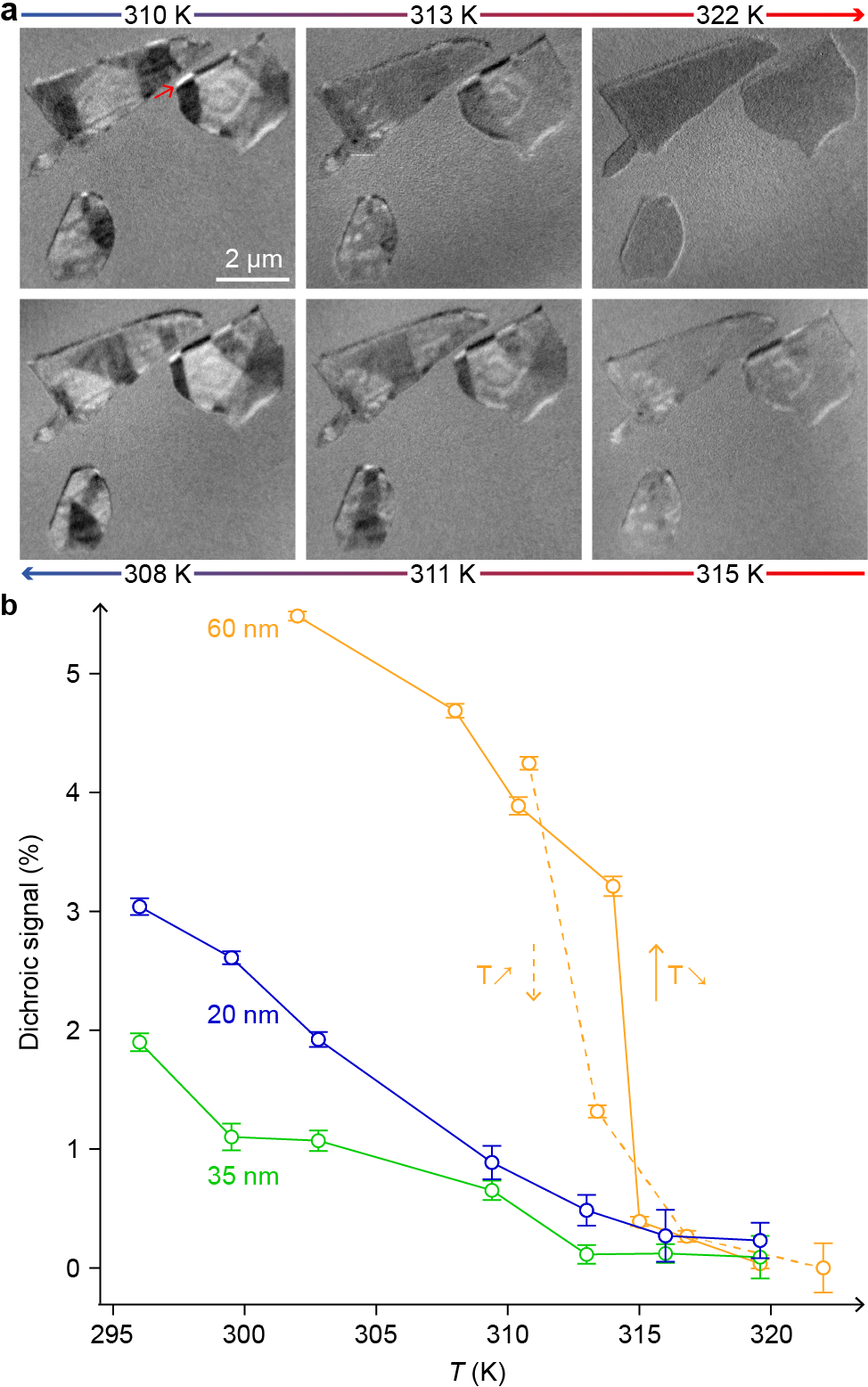}
 \caption{\label{fig4} \textbf{Curie temperature of individual flakes.} (a) XMCD-PEEM magnetic images (576.6~eV) of the same flakes for different sample temperatures (the three flakes are about 60~nm-thick). The top row shows the warming process, whereas the bottom row shows the cooling. The red arrow shows the position of the flake shadow, which also carries a magnetic contrast. (b) Intensity of the magnetic contrast deduced from (a) and from another set of images (not shown) allowing an estimate of the Curie temperature. Note that the contrast varies from one data set to the other (green-blue-orange) depending, among other factors, on the sample preparation conditions (time before introduction under ultrahigh vacuum) and settings of the XMCD-PEEM instrument.}
 \end{center}
\end{figure}

The XMCD-PEEM images reveal flux-closure configurations (see also Figure~\ref{fig4}), and magnetic domains often compatible with a six-fold symmetry (see Figures~\ref{fig3}a, \ref{fig4}). Consistent with this observation, the walls between the domains often orient (not exclusively, though) according to a six-fold symmetry (Figure~S4), as shown by a statistical analysis (Figure~S5) -- they are 120$^\circ$ magnetic domain walls.
These observations suggest the presence of a nonzero in-plane magnetocrystalline anisotropy, although this anisotropy might be small, since six-fold symmetry is not systematically observed (Figure~S5). This points to a contribution of shape anisotropy, which promotes other preferential in-plane orientations given that the flakes rarely exhibit edges corresponding to specific crystallographic directions in CrTe$_2$ (Figure~S5; such directions would probably be distributed according to a six-fold symmetry). A six-fold magnetocrystalline anisotropy is consistent with the structure of 1T-CrTe$_2$, a trigonal system with hexagonal unit cell.

Such six-fold anisotropy has been observed in a few thin film systems, for instance made of copper-cobalt superlattices,\cite{Schreiber} iron-cobalt-platinum,\cite{Nahid} or cobalt-implanted zinc oxide,\cite{Akdougan} but not for van der Waals ferromagnets to our knowledge. In fact, for another van der Waals ferromagnet, K$_2$CuF$_4$, a preferred in-plane orientation was predicted with no six-fold in-plane anisotropy.\cite{Sachs}

Across domain walls the MFM phase signal, which is proportional to the gradient of the force experienced by the MFM tip, first increases and then decreases (or \textit{vice versa}) above and below the constant value corresponding to the two domains (see the profiles measured perpendicular to the walls in Figure \ref{fig3}c). In other words, the phase signal is a function of the position having a strong odd symmetry character about the center of the domain wall. Within the common description relying on bond charges (giving rise to the magnetization) of magnetostatics, such odd symmetry character is a clearcut qualitative evidence of an accumulation of positive and negative charges on either sides of the domain wall.\cite{Hartmann}

Can we tell more about the texture of the magnetisation within the domain walls? In principle, several kinds of textures could account for our observations, unfortunately very few experimental techniques (if any) would be able to directly resolve them. As we discussed, for CrTe$_2$ flakes we observe 120$^\circ$ in-plane magnetised domains. In such a case, the texture of the domain walls is expected to be of N\'{e}el type, and not of Bloch type,\cite{Hubert} \textit{i.e.} the magnetisation should rotate within the surface plane all along the thickness of the domain walls. In the thickness range that we probed, and given the relatively small $M_\mathrm{S}$, the N\'{e}el walls should in fact be symmetric.\cite{Hubert} These expectations are not very sensitive to the anisotropy constant value,\cite{Hubert} which is unknown so far for CrTe$_2$. The symmetric N\'eel texture is associated with an accumulation of positive and negative charges on either sides of the wall, and is thus fully consistent with our MFM observations.

The XMCD-PEEM images indicate that the domain wall width is of the order of several tens of nanometers (see intensity scans in Figures~\ref{fig3}d,e). 
Finally, combining the different information provided by the two techniques, the local magnetization texture can be resolved, including that of the magnetic domain walls (see arrows in Figure~\ref{fig3}a).

The thinnest flake we imaged (10 nm) does not show any MFM or dichroic magnetic contrast, and no hysteresis loop is measured (see Figure \ref{fig2}). This observation is not a clear-cut evidence for T$_\text{C} <$ 300~K in this case, though, and could be instead interpreted by invoking the limited sensitivity of the experimental probes.
The 100- and 30-nm-thick flakes (see Figure~\ref{fig3}a) can be studied using both imaging techniques and with local magnetometry as well. 
However, the 80-nm-thick flake (see Figure~\ref{fig3}b) is characterized by an unambiguous hysteresis loop and a faint MFM contrast, but no clear XMCD-PEEM magnetic contrast within the flake.
This finding suggests that the top of this particular flake might be degraded over a thickness exceeding the photoelectrons' mean-free path of the order of a nanometer, preventing magnetization to be probed with a surface sensitive technique such as PEEM, even if the flake is in fact ferromagnetic at room temperature. 
This surface degradation is also compatible with the analysis of the AFM phase contrast which, together with Raman spectroscopy and x-ray absorption spectroscopy, reveals that this region of the flake might be oxidized (see Figure~S6).
We note that the extent of the surface degradation we observe is flake dependent and does not seem to be related to its thickness.

\subsection*{Ferromagnetic-paramagnetic transition} XMCD-PEEM imaging was also used upon sample annealing to track the temperature dependence of the magnetic contrast, and to locally measure the Curie temperature T$_\mathrm{C}$. 
As expected, the magnetic contrast continuously drops as the temperature increases (see Figure~\ref{fig4}a). 
This behavior is observed for several CrTe$_2$ flakes of different thicknesses (the thickness of each flake was determined \textit{a posteriori} with AFM).
For all the imaged flakes, the magnetic contrast vanishes around 315~K, which is also the T$_\mathrm{C}$ value of the bulk material (see Figure~S3).\cite{Freitas2015} 
Our results do not reveal any obvious change in T$_\mathrm{C}$ when the thickness of the CrTe$_2$ flakes is varied, at least in the range we probe in this work, for which no thickness dependency has been observed either for other van der Waals ferromagnets.\cite{Gibertini2019}
Our measurements also show that the magnetic domain structure may vary upon cycling the temperature above the Curie point, indicating that there is no significant magnetic memory of the initial state.  

Interestingly, because of the large thickness of the flakes and the grazing incidence of the x-ray beam in XMCD-PEEM measurements, a shadow is projected onto the substrate surface (see red arrow in Figure~\ref{fig4}a). 
The dichroic signal in the transmitted intensity seen within the shadow is identically inverse of the dichroic contrast due to the direct emission from the CrTe$_2$ surface\cite{Kimling2011,DaCol2014,Jamet2015}. 
We thus conclude that the magnetization is homogeneous within the bulk of the flake, which is reliably represented by the XMCD-PEEM image obtained from the surface.

\begin{figure}[!hbt]
 \begin{center}
 \includegraphics[width=16cm]{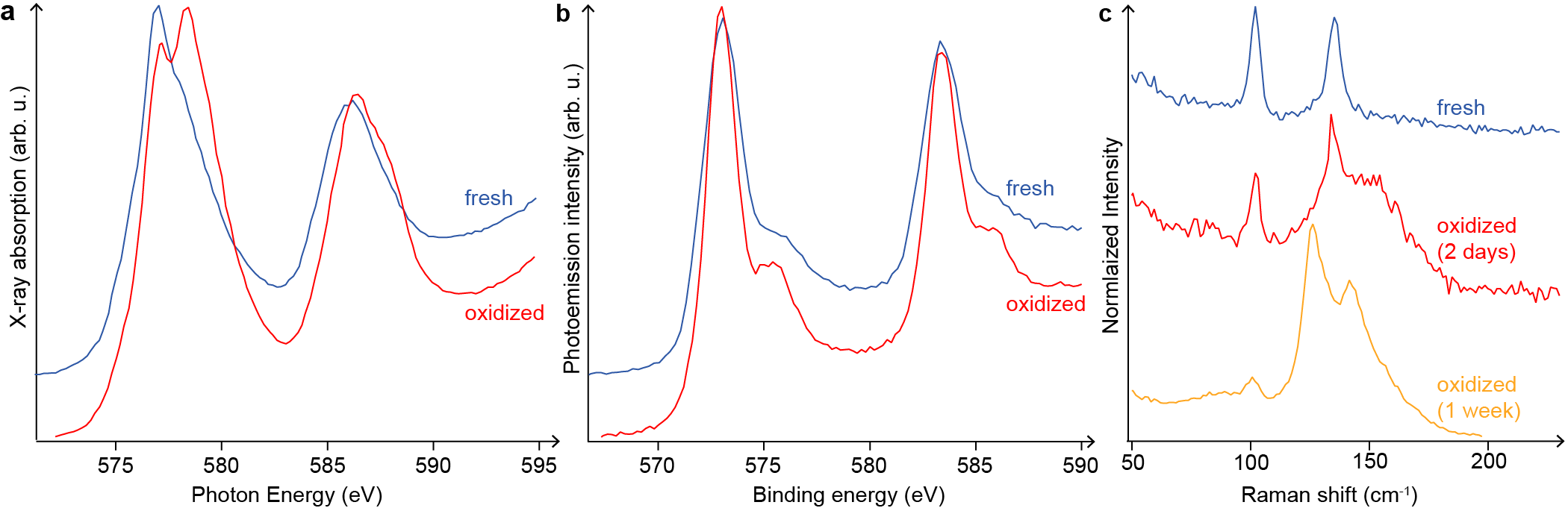}
 \caption{\label{fig5} \textbf{Sample aging and formation of an oxide layer.} (a) Absorption spectra at the L$_3$ and L$_2$ Cr edges for a 20~nm-thick freshly exfoliated flake (blue curve) and after one day under ambient conditions (red curve). (b) X-ray photoemission spectroscopy (Cr 2p$_{3/2,1/2}$ core levels) of the same sample before (blue curve) and after (red curve) exposure to ambient conditions. (c) Raman spectra of the sample as a function of time.}
 \end{center}
\end{figure}

\subsection*{Oxidation in ambient conditions} All magnetic vdW materials known to date are reactive materials, and often require cautious manipulation.\cite{Deng2018,Shcherbakov2018} 
We thus discuss here the stability of the CrTe$_2$ compound. 
The magnetic contrast observed in XMCD-PEEM shows a reduced dichroic signal for samples exposed to ambient conditions: the contrast is typically reduced by one order of magnitude between a freshly cleaved sample and a sample exposed to air for several days.
Considering the mean-free path of the photoelectrons used for imaging, this observation suggests that CrTe$_2$ oxidizes with time upon exposure to ambient conditions, and that the oxidation is limited to the outer layers (typically 1-2~nm) of the flakes.
X-ray absorption spectroscopy (XAS) and photoelectron spectroscopy (XPS) measurements were performed as a function of ambient exposure time, starting from freshly prepared CrTe$_2$ flakes (see Figure~\ref{fig5}).
The XAS spectra progressively evolve, revealing a multiplet-like Cr L$_{2,3}$ absorption edges for samples exposed to air, whereas they are essentially metallic on freshly cleaved flakes (see Figure~\ref{fig5}a), with binding energies close to the values obtained for clean, pure Cr.\cite{Schwickert1998} 
Consistently, while Cr core level 2p$_{3/2,1/2}$ binding energies (573.0, 583.4~eV) of freshly prepared flakes are close to those of pure Cr surfaces,\cite{Allen1978} additional peaks are observed at higher binding energies (+2.4~eV) after prolonged air exposure (see Figure~\ref{fig5}b).
We attribute the presence of these peaks to the formation of a chromium oxide layer.\cite{Allen1978}

Contrary to the x-ray measurements which require to expose the sample to air for about 30~min, Raman spectroscopy can be performed under the ultra-clean environment of an Ar glove box (including the exfoliation). Such freshly prepared flakes were sealed \textit{in situ} inside a quartz tube, and Raman spectra were recorded on the sealed flakes, \textit{i.e.} prior to any exposure to air. No noticeable difference was observed between the Raman spectra measured for flakes exposed to air for about 30~min (Figure~\ref{fig1}c,\ref{fig3}c) and for sealed flakes.

The two Raman peaks (E$_{2g}$, A$_{1g}$) of freshly exfoliated or sealed flakes, observed at 102 and 134~cm$^{-1}$, are replaced after one weak of exposure to air by two other peaks, at 126 and 142~cm$^{-1}$ (see Figure~\ref{fig5}c). In an intermediate degradation state, the peak at 126~cm$^{-1}$ essentially appears as a shoulder of the initial 134~cm$^{-1}$ peak, and the one at 142~cm$^{-1}$ as a much broader peak (see Figure~\ref{fig5}c).

The $\gamma$ phase of TeO$_2$ (obtained after re-crystalisation of TeO$_2$ glass) exhibits two Raman peaks at 116 and 140~cm$^{-1}$, that are close to those we observe.\cite{ChamparnaudMesjard2000} The corresponding vibration modes were described with the A1 irreducible representation.\cite{ChamparnaudMesjard2000} In a much related phase of TeO$_2$, the $\alpha$ phase, the two modes are upshifted by $\sim$5 and $\sim$10~cm$^{-1}$ respectively.\cite{ChamparnaudMesjard2000}

The glassy nature of TeO$_2$ suggests, in our case where no special thermal treatment has been performed, the coexistence of various phases with more-or-less defined structure, and each characterised by Raman modes of slightly different intensity. This might account for the broad peak observed around 140-150~cm$^{-1}$ at an intermediate state of the oxidation process. Time being a key factor in a kinetically-limited oxidation process such as here (occurring at room temperature), we propose that the corresponding peak becomes narrower with time, eventually leading to the (relatively) narrow peak centred at 142~cm$^{-1}$ corresponding to one specific (more-or-less crystallized) TeO$_2$ phase.

\section*{Conclusions}

Combining several spatially-resolved techniques sensitive to magnetism, room temperature ferromagnetism is demonstrated in exfoliated CrTe$_2$ flakes of thickness down to 20~nm. 
The Curie temperature of 315~K is similar to that of the parent bulk compound, whereas the weak coercivity measured experimentally shows that the material is magnetically soft. These results are consistent with those reported in Ref.~\citenum{Sun2020} using magneto-transport measurements.
Magnetic imaging reveals the magnetic domain microstructure. 
In particular, large in-plane magnetic domains are observed, forming flux-closure configurations with a tendency to six-fold symmetry. Such patterns explain the absence of remanence observed with ensemble-averaging probes used to address bulk samples, while our microscopic analysis reveals locally almost-100\% remanence. The magnetic domains are separated by several 10~nm wide domain walls presumably of N\'eel type.
These findings suggest that the magnetic patterns in the CrTe$_2$ flakes results from both shape and magnetocrystalline anisotropies. 
This new thin van der Waals ferromagnet is moderately reactive to oxidation under ambient conditions, with the topmost few layers of the material being oxidized typically within a day.

Our work introduces a new member in the growing family of van der Waals ferromagnets, with in-plane easy axis and a Curie point above room temperature, which is attractive for both applications and fundamental studies. 
In the prospect of applications, an in-plane magnetized room-temperature ferromagnet is attractive to build heterostructures hosting specific quantum phases, such as an anomalous quantum Hall effect,\cite{Liu2013} or artificial antiferromagnets with two CrTe$_2$ layers of different thicknesses. 
Concerning fundamental studies, thinning the CrTe$_2$ flakes down to the single layer limit is an exciting goal. This calls for an optimization of exfoliation processes, beyond what is already established for more traditional materials like graphite and MoS$_2$.
In addition, the use of highly sensitive magnetic probes,\cite{Thiel2019} should allow to determine whether the Curie temperature of the compound decreases with the layer thickness. This would tell us whether the absence of magnetic contrast that we found for 10~nm-thin CrTe$_2$ flakes is related to the limited sensitivity of MFM and XMCD-PEEM, or is rather intrinsic to the material itself, \textit{i.e.} a manifestation of a truly two-dimensional character.

\section*{Materials and Methods}

\textbf{Bulk CrTe$_2$ synthesis.}
1T-CrTe$_2$ samples were synthesized indirectly by oxidation of KCrTe$_2$. 
The parent compounds were prepared by a molar mixture of the elements, Cr, K and Te under argon atmosphere in a glove box.
This mixture was heated to the melting point of both the alkali element and tellurium, and then kept at 900$^\circ$C for eight days in an evacuated quartz tube. 
The tubes were opened in a glove box in order to prevent oxidation. 
Alkali atom de-intercalation was then carried out by reacting the parent compounds in solutions of iodine in acetonitrile. 
These suspensions were stirred for about 1 hour, using an excess of iodine. 
The final product was washed with acetonitrile to remove the formed iodide, filtered and dried under vacuum. A mixture of brilliant gray platelets about 3 mm-long, 0.3 mm-thick were obtained.

\noindent \textbf{Magneto-optical focused Kerr magnetometry.}
A He-Ne laser (wavelength 632 nm) beam of 5~mW power was linearly polarized (s-polarization) and a 100$\times$ objective lens focused the laser beam to a spot with a diameter of less than 1~$\mu$m on the sample. 
The reflected signal was separated by a Wollaston prism into two beams with orthogonal polarizations impinging on two identical photodiodes. 
The sum and difference signal of the two diodes was sent to an oscilloscope. 
The sample was placed in the gap of a small horseshoe magnet that allowed applying an in-plane magnetic field of up to 80 mT.
The signal on the oscilloscope was measured during a field sweep at a frequency of 1 or 2~Hz with a maximum field of $\pm$10~mT.
To improve the signal to noise ratio, the hysteresis loops were averaged over about 100 field sweeps.

\noindent \textbf{Magnetic force microscopy.}
Magnetic images were obtained using a NT-MDT microscope. 
The magnetic layers of the scanning tips were Co$_{80}$Cr$_{20}$, a semi-hard material with coercivity of the order of 100 mT. This material was used to cap (50-75~nm thickness) our home-made MFM tips, whose magnetisation is essentially perpendicular to the surface. Like usually the case in MFM experiments, we cannot provide more details about the magnetisation configuration of the tip, hence we cannot extract quantitative information about the sample magnetisation. For the second pass image, acquired after the topographic one a tip lift height of about 100~nm yielded optimum MFM phase contrast and resolution, and the domain structure of the flakes seemed unaffected by the scanning with the MFM magnetic tip (our observation with MFM and XMCD-PEEM are fully consistent).

\noindent \textbf{X-ray photoemission electron microscopy.}
The XAS, XMCD and XPS-PEEM measurements were performed using the SPELEEM III microscopes (Elmitec GmbH) installed at the Elettra\cite{Elettra} and ALBA\cite{ALBA} facilities. 
The lateral resolution in XAS-PEEM nears 20~nm for both microscopes. 
The XMCD-PEEM images shown here were obtained by subtracting PEEM images acquired with opposite helicities of the incoming photon beam and normalized to the sum of these images. 
Magnetic images are acquired at the Cr L$_3$ absorption edge.

\noindent \textbf{X-ray absorption and photoelectron spectroscopy.}
XAS and XPS spectra were obtained from scans of either the photon energy (XAS) or the detected electron energy (E$_{kin}$ or start voltage). 
For each step, a PEEM image is acquired and local (XAS,XPS) spectra are extracted by plotting the averaged intensity in a selected area vs the scanned parameter. 
For XPS, the binding energy was obtained as $E_{bind}$ = $E_{photon}$ - $E_{kin}$ - $\phi_{analyzer}$.   

\noindent \textbf{Raman spectroscopy measurements.} 
The Raman spectra of the CrTe$_2$ flakes were recorded using a Witec Alpha 500 Raman microscope with Rayshield coupler, which is able to acquire low frequency spectra as low as 10  rel.cm$^{-1}$. 
A laser with an excitation wavelength of 532 nm was focused to the sample with a 50$\times$ Mitutoyo, NA 0.42 objective. 
The power of laser excitation measured below the microscope objective lens was set to 0.1 mW to avoid sample heating. 
A 1800 lines/mm grating was set to achieve higher resolution, $\leq$ 0.1 cm$^{-1}$. 
The spectra were collected with integration times of 30~s and number of accumulation of 5. All measurements were performed at room temperature.

%\begin{suppinfo}
%Supporting Information includes \johann{...}.
%\end{suppinfo}

\section{Author Information}
*E-mail: johann.coraux@neel.cnr.fr

\section{Associated Content}
The authors declare no competing financial interest.

\begin{acknowledgement}

This work was financially supported by the Flag-ERA JTC 2017 project `MORE-MXenes'.
A.P. and N.R warmly thank Simon Le-Denmat and Olivier Fruchart for technical help during the MFM measurements, and Michel Hehn for providing the Pt/Ta/Si substrates.

\end{acknowledgement}

\begin{suppinfo}

Supporting information comprises supplementary Raman spectroscopy data as function of light polarization and flake thickness, XMCD-PEEM analysis of magnetic anisotropy, AFM and Raman spectroscopy analysis of the flakes' oxidation, and an analysis of the magnetization versus temperature evolution in thick samples.
\end{suppinfo}

%\bibliography{CrTe2}

\begin{mcitethebibliography}{58}
\bibitem[not()]{note1}
In the 1T polytype of a transition metal dichalcogenide, the local environment
  of the metal atoms is octahedral, and each atomic position in an individual
  layer sits directly atop the corresponding site in the next layer. Altogether
  this yields a threefold symmetry axis perpendicular to the layers,
  characteristic of a trigonal crystal system.\relax
\mciteBstWouldAddEndPunctfalse
\mciteSetBstMidEndSepPunct{\mcitedefaultmidpunct}
{}{\mcitedefaultseppunct}\relax
\EndOfBibitem
\bibitem[not()]{noteondust}
Three clusters enclosed in the blue-circled area. These clusters are ascribed
  as dust particles or glue residues, as discussed in the first section. They
  represent about 10\% of the blue-circled area, and a height variation of
  10~nm. Given the height of the flake (80~nm), the clusters overall represent
  slightly more than 1\% of the total volume probed in the experiment. We thus
  consider their contribution to the MFM phase signal as negligible.\relax
\mciteBstWouldAddEndPunctfalse
\mciteSetBstMidEndSepPunct{\mcitedefaultmidpunct}
{}{\mcitedefaultseppunct}\relax
\EndOfBibitem
\end{mcitethebibliography}

\providecommand{\latin}[1]{#1}
\makeatletter
\providecommand{\doi}
  {\begingroup\let\do\@makeother\dospecials
  \catcode`\{=1 \catcode`\}=2 \doi@aux}
\providecommand{\doi@aux}[1]{\endgroup\texttt{#1}}
\makeatother
\providecommand*\mcitethebibliography{\thebibliography}
\csname @ifundefined\endcsname{endmcitethebibliography}
  {\let\endmcitethebibliography\endthebibliography}{}

\end{document}